# Supersonic Impact of Metallic Micro-particles

Mostafa Hassani-Gangaraj[1], David Veysset[2,3], Keith A. Nelson[2,3], Christopher A. Schuh[1]*

[1]Department of Materials Science and Engineering, MIT, Cambridge, Massachusetts 02139, USA.

[2]Institute for Solider Nanotechnologies, MIT, Cambridge, Massachusetts 02139, USA.

[3]Department of Chemistry, MIT, Cambridge, Massachusetts 02139, USA.

*Correspondence to: schuh@mit.edu

**Understanding material behavior under high velocity impact is the key to addressing a variety of fundamental questions in areas ranging from asteroid strikes[1,2] and geological cratering[3] to impact-induced phase transformations[4], spallation[5], wear[6], and ballistic penetration[7]. Recently, adhesion has emerged in this spectrum since it has been found that micrometer-sized metallic particles can bond to metallic substrates under supersonic-impact conditions[8,9]. However, the mechanistic aspects of impact-induced adhesion are still unresolved. Here we study supersonic impact of individual metallic microparticles on substrates with micro-scale and nanosecond-level resolution. This permits the first direct observation of a material-dependent threshold velocity, above which the particle undergoes impact-induced material ejection and adheres to the substrate. Our finite element simulations reveal that prevailing theories of impact-induced shear localization and melting cannot account for the material ejection. Rather, it originates from the propagation of a pressure wave induced upon impact. The experiments and simulations together establish that the critical adhesion velocity for supersonic microparticles is proportional to the bulk speed of sound.**

The phenomenon of impact-induced adhesion of microparticles has presented many interesting fundamental questions about materials response under extreme conditions. It has also opened a new window in additive manufacturing, as an alternative to high-temperature fusion or sintering of particles[10,11] to produce bulk component; many metallic particles can be accelerated through supersonic nozzles and impacted onto metallic substrates to build a solid material[12,13], even when there is no applied heat to inspire, e.g., melt formation or liquid-phase bonding. In this area, researchers have repeatedly observed a "critical velocity", a threshold above which supersonic particles adhere to the substrate instead of rebounding[14–16]. A variety of mechanisms have been put forth to explain this empirical observation, such as adiabatic shear instability[8], localized melting[17], viscous type mechanical interlocking[18], interface amorphization[19], and oxide-layer break-up[20]. However, these putative mechanisms have not been quantitatively supported by physical theories nor yet directly observed. In our view, the lack of consensus on the operative mechanisms traces to a lack of real-time studies of supersonic micro-particle impact. Such studies require spatial (micron) and temporal (nanosecond) resolutions much finer than those provided by existing experimental techniques.

Here we conduct the first in-situ single-particle study of supersonic microparticle adhesion. We employ an in-house-designed microscale ballistic test platform[21–23] to accelerate micrometer-size metallic particles and observe the entire deformation/adhesion process in real time. As schematically shown in Fig. 1a, a laser excitation pulse is focused onto a launching pad assembly from which single metallic particles are launched toward a target sample by ablation of a gold layer and rapid expansion of an elastomeric polyurea film. The particle approach and impact on the target are observed in real time using a high-frame-rate camera and a synchronized quasi-cw laser imaging pulse for illumination, such as described in ref[23].

Our method resolves the instant of impact with micrometer-scale spatial resolution and nanosecond-level temporal resolution. Fig. 1b shows some exemplar results taken for 45-µm Al particles impacting an Al target with velocities slightly below and above the critical velocity: 605 and 805 m/s (± 4%) respectively. At sub-critical impact velocity, the particle rebounded with clear flattening and considerable deformation. At above-critical velocity, the particle did not rebound but instead adhered to the substrate. These images, also montaged into supplementary videos 1 and 2, are the first direct observations of the rebound-adhesion transition. Processing images such as those in Fig. 1b (see Methods) leads to data such as shown in Fig. 1c, where the coefficient of restitution, defined as the ratio between the rebound velocity and impact velocity, is shown as a function of impact velocity for four materials. As the impact velocity increases, the coefficient of restitution decreases until it eventually goes to zero, revealing the critical velocity for adhesion; the method has enabled the first direct measurements of critical adhesion velocity for Cu, Ni, Al and Zn particles to matched-material substrates.

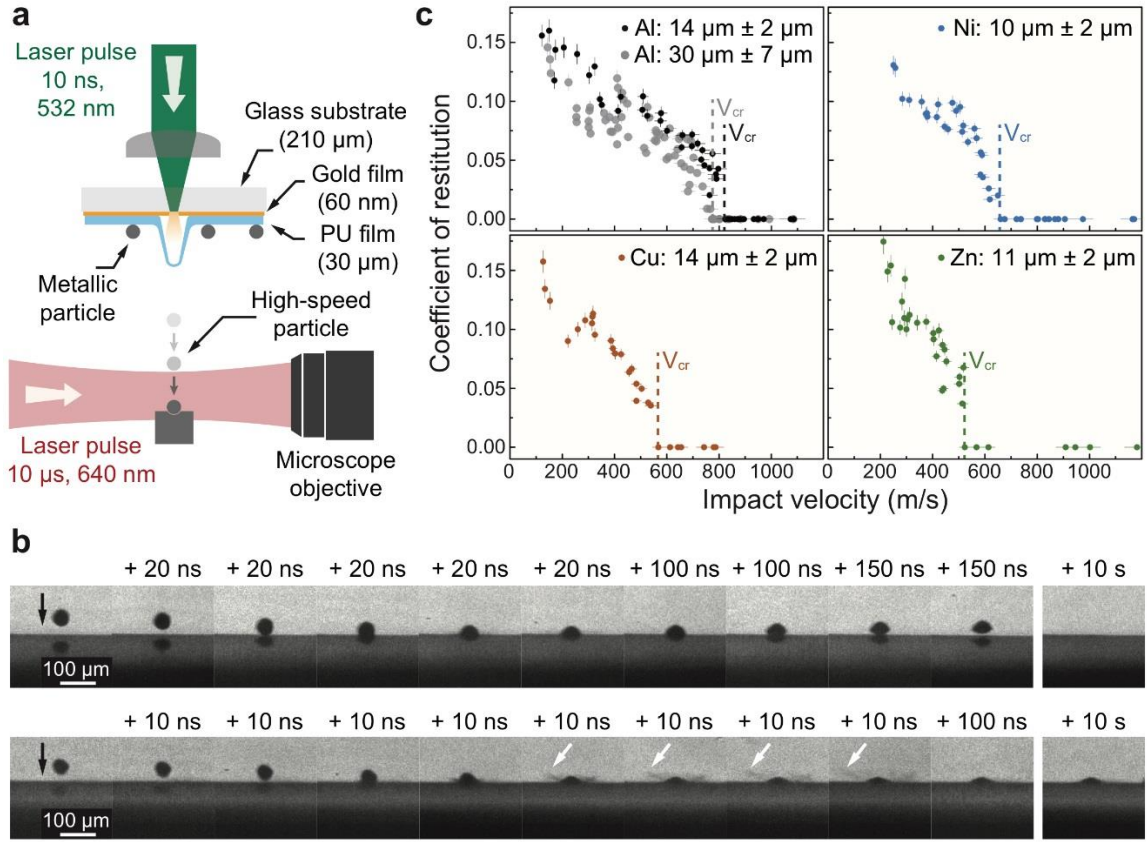

**Figure 1 | In situ observation of microparticle supersonic impact. a**, Experimental platform for microparticle impact test and real-time imaging. **b**, Multi-frame sequences with 5 ns exposure times showing 45-µm Al particle impacts on Al substrate at 605 m/s (top) and 805 m/s (bottom), respectively below and above critical velocity. The micro-projectiles arrive from the top of the field of view. Material jetting is indicated with white arrows. **c**, Coefficient of restitution for Al, Ni, Cu, and Zn. The coefficient of restitution is equal to zero above the critical velocity.

Prior to this work, the critical velocity has been determined based on deposition efficiency measurements in cold spray experiments[16]. In such experiments the complex interactions of the particles with the hot carrier gas, the bow shock effect in front of the substrate, the lack of knowledge of individual particles' velocity and the indirect nature of the measurement all confuse the situation, compromising the accuracy and cleanliness of the data. Conversely, in our data in Fig. 1 we observe a clear discontinuity in the coefficient of restitution at the critical velocity, following its roughly linear decline at lower velocities. In addition to providing a clean measurement of the critical velocity, the discontinuity also suggests the emergence of a physical phenomenon governing particle deformation at critical and above-critical impact velocities. Greater insight is provided by the images in Fig. 1b, where the above-critical particle images show significant plastic deformation of the particle accompanied by very fast lateral material jet-like ejection at the periphery of the particle.

Based on the above observations, we hypothesize that the formation of an interfacial jet and plastic ejection of material is critical to supersonic adhesion. Large plastic deformation, caused by solid-state jetting, provides fresh metallic surfaces and facilitates pristine atomic contact between particle and substrate leading to bonding, in a similar manner as for explosive welding[24]. The same bonding process has been suggested to apply for cold spray coating[8,9] but has not been directly supported by observations at the individual particle level on the time scale of the impact event itself. To evaluate this hypothesis on the basis of deformation physics, we simulated particle behavior upon impact using a 3D coupled thermo-mechanical dynamic finite element model (see Methods). The main constituents of material behavior in the model include hydrodynamic response[25] and strain-, deformation rate- and temperature-dependent plasticity[26]. Importantly, the method used here permits fragmentation and surface creation without needing explicit interface-tracking, and can therefore capture jetting and material ejection when it is mechanically favored. Images taken from finite element simulations of copper (Fig. 2a) show that spherical particles deform smoothly into an oblate shape at sub-critical impact velocities, with no sign of jet formation despite relatively large deformation. Past a threshold velocity, however, our simulations (Fig. 2b) feature lateral material ejection, formation of a material jet, and eventual fragmentations in the jet region. The particle deformation features shown in Fig. 2 are exemplary of many that we have produced, and align well with the direct experimental observations in Fig. 1. The agreement between the measured critical velocities and the velocities corresponding to the onset of jetting in simulations, shown in Fig. 2c, along with real time observation of jetting before adhesion (Fig 1b) and postmortem images of the adhered particles, shown in Extended Data Fig. 1, together all suggest that jet formation and subsequent material fragmentation is triggered at the critical impact velocity, and that this is directly associated with adhesion.

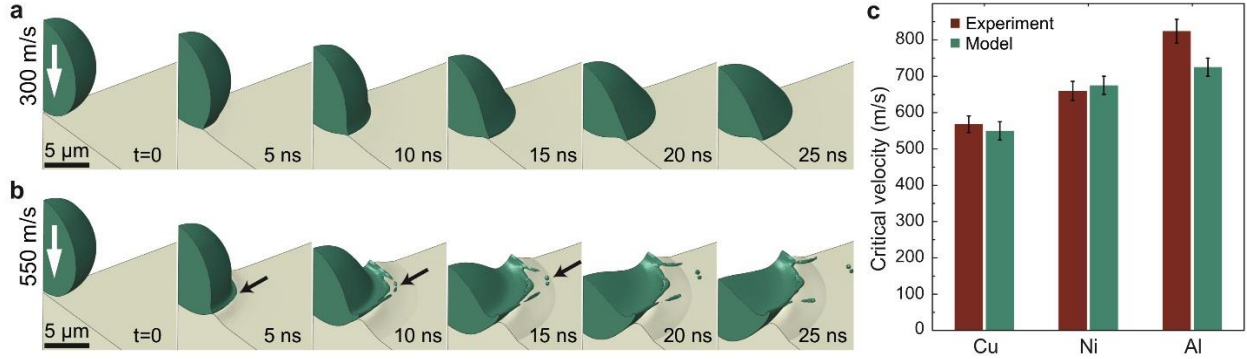

**Figure 2 | Computer simulations of microparticle impact**. Deformation of a 10-μm Cu particle at sub-critical (**a**) and critical (**b**) impact velocities. Particles impacted at 300 m/s and 550 m/s reach their maximum substrate penetration after 20 and 25 ns respectively. Formation of an interfacial jet and plastic ejection of material (shown with black arrows) occur only when the impact velocity exceeds a material-dependent threshold. **c**, Comparison of experiments and model. The small differences between the experiments and model could arise from potential particle shape deviations from ideal spheres, or the slight particle size differences between the experiments and model and the uncertainty in material constitutive parameters that are determined from bulk forms.

These simulations can be used to further explore the underlying physics of material jet formation. A first observation is that local melting is not likely involved: Extended Data Fig. 2 shows that even at its peak, the plastic-work-induced temperature rise reaches far less than the melting temperature for any of the materials tested. This rules out localized melting and consequent viscous flow as the origin of material jet formation. Our simulations also suggest that adiabatic shear localization is not likely a cause of the jetting: the required thermal softening would cause the material strength to fall to values close zero[27], whereas Extended Data Fig. 3 shows that the yield strength is decidedly not compromised when the material disintegration first occurs in an already-formed jet. Adiabatic localization is thus not the cause of material jet formation either, although it may be a consequence and occur as the jet expands further.

Classical studies of liquid drop impact on solid and liquid surfaces point to the interaction of a shock wave with the particle leading edge as the origin of jetting[28,29]. When the shock detaches from the contact edge and as it moves up the free surface of the particle, release waves propagate from the contact point and accelerate material to a high speed, forming a jet. It is clear from our simulations as shown in Fig. 3 that this mechanism is at play in supersonic solid particle impact. A strong compressive shock is generated upon impact at the critical velocity and first remains attached to the leading edge as shown by the first two images. Within 2 ns after the initial contact and as the contact zone expands, the compressive shock overtakes the leading edge. As a result, release waves induce material jetting in that region as shown by the image at 2.1 ns. The jet further develops until it fragments, as evidenced by the image at 9.1 ns.

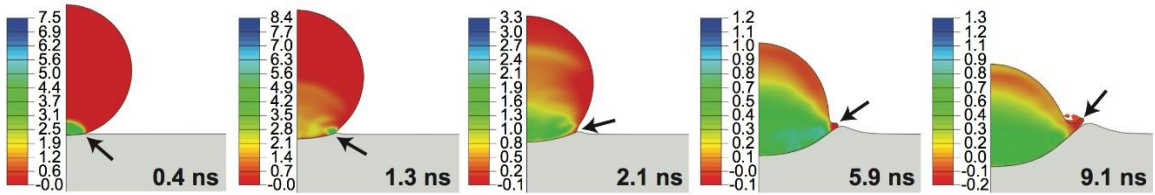

**Figure 3 | Simulations showing the impact-induced pressure during a critical velocity (550 m/s) impact of a 10-μm Cu particle.** The values are expressed as percentage of the bulk modulus. Positive values indicate hydrostatic compression. At the early stage of deformation, the pressure is on the order of 10% of the bulk modulus, and the interaction of the compressive shock with the free surface is shown to be the origin of material jet formation and fragmentation.

Because jetting is a pressure-governed mechanism and the pressure generated upon impact is directly related to the impact velocity through the bulk speed of sound (Eq. 1 in SI), we expect the bulk speed of sound to be the controlling parameter in adhesion. Indeed, when comparing across the different materials we have tested, we find a linear relation between the critical velocity and the bulk speed of sound (Fig. 4). These results can be summarized with a simple predictor for the critical impact velocity for adhesion of ~10 μm particles:

$$V_{cr,ref} = \alpha \sqrt{\frac{B}{\rho}} \tag{1}$$

where $B$ is the bulk modulus and $\rho$ is density, the square root of their ratio is the speed of sound, and $\alpha$ is a constant. $\alpha = 0.15$ is suggested by fitting the experiments, while a similar value of $\alpha = 0.14$ is found by fitting the simulations.

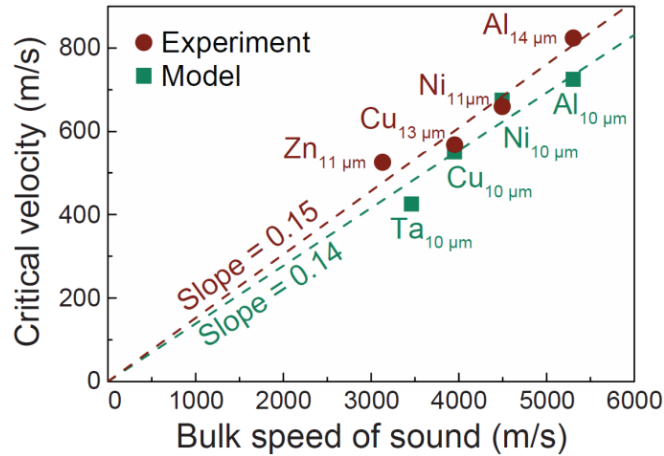

**Fig. 4 | Revealing critical velocity relation with bulk speed of sound.** Together our experiments and simulations reveal a linear relation between critical adhesion velocity and the bulk speed of sound in metallic materials. The slope for ~10-μm particles impacting matched material substrates at room temperature is about 0.15.

We are not aware of prior predictions of critical velocity that speak to the physics of impact-induced adhesion, i.e. high pressures and jetting governed by the bulk speed of sound. The existing relations used to predict critical velocity[8,13] are only empirically calibrated and not mechanistic. What is more, they suggest a dependency on materials strength, which our analysis and data show is in fact not physically appropriate; under identical experimental conditions, softer Al requires higher impact velocity than harder Ni to adhere. This is a reasonable result since under supersonic impact conditions the stress levels produced swamp the material strength level by orders of magnitude. To further confirm this, our simulations (Extended Data Fig. 4) show negligible effect of material strength until it approaches a significant fraction of the impact-induced pressures.

Eq. 1 is a simple but powerful heuristic, and it can of course be broadened to incorporate many additional physically relevant terms. For example, in the Supplementary Information we show how it can be semi-empirically expanded to include the particle initial temperature (T) and size (d) effect on the critical impact velocity. Increasing particle temperature leads to greater flattening upon impact, and decreases the critical velocity in a manner that follows a square root dependency on the dimensionless temperature, $T^* = (T-T_{room})/(T_{melt}-T_{room})$ (Extended Data Fig. 5). Increasing particle size results in higher temperatures in the jet region, which in turn affects critical velocity (Extended Data Fig. 6). The size effect described in the Supplementary Information is verified by the impact experiments on larger Al particles shown in Figure 1b, which confirms a shift in critical velocity from 825 m/s for 14 micron particles to 775 m/s for 30 micron Al particles, agreeing well with our modeling.

To summarize, our nanosecond and microscale in-situ observations of impact adhesion have provided much needed mechanistic insight to this extreme-conditions phenomenon. By offering contrary viewpoints to some widely postulated mechanisms including adiabatic shear instability or melting, we propose a new direction of focus: compressive shock interaction with the particle leading edge results in material jetting and a mixing flow that provides the clean metallic contacts essential for adhesion at the particle/substrate interface. The simple relation between critical velocity and the bulk speed of sound established by Eq. 1 and its embellishment with additional physical factors should prove useful for the understanding of impact-induced adhesion across a range of materials, and for the design of additive manufacturing processes that rely on impact adhesion.

## METHODS

**Sample Preparation.** Two batches of Al powder particles with nominal particle sizes of 20 and 31 µm were purchased from Valimet (Stockton, USA). Cu, Ni, and Zn with nominal particle sizes of 10 µm, 5-15 µm and 6-9 µm respectively, were purchased from Alfa Aesar (Ward Hill, USA). Al, Zn, Ni and Cu plates with 3.175 mm thickness were purchased from OnlineMetals (Seattle, USA). We used a water jet cutter to extract 15×15×3.175 mm plates for use as the targets for the impact experiments. Each target surface was ground and polished to 1 µm mirror finish prior to the impact experiments.

**Launching Pad Assembly.** The launching pad assembly follows the design described by Lee et al. and Veysset et al. in refs[22,23]. 10-µm-thick glass substrates (Corning No. 2 microscope cover slip) were sputter-coated with a 60-nm thick gold film. A mixture of polycarbodiimide-modified diphenylmethane diisocyanate (Isonate 143L MDI, Dow Chemicals) and oligomeric diamine (Versalink® P-650, Air Products) with a weight ratio of 1:3 was spin-coated on the gold-coated substrates at 750 RPM for 5 min to yield a film thickness of 30 µm after 24-hour curing at room temperature. Film thicknesses were measured using a 3D laser scanning confocal microscope (VK-X200 series, Keyence). Metallic particles were deposited on the substrates using lens cleaning papers to spread drops from a suspension of particles in ethanol.

**Impact Experiments.** For each experiment, a laser excitation pulse, 10-ns in duration and using 532-nm wavelength light (Pulsed Nd:YAG, Quanta-Ray, Spectra Physics) was focused onto the launching pad assembly from where the metallic particles were ejected. Upon laser ablation of the gold film, particles were accelerated to speeds ranging from approximately 100 to 1200 m/s, controllable by adjusting the laser excitation pulse energy (from 2 up to 60 mJ). 16-image sequences showing impact were recorded with a high frame rate camera (SIMX 16, Specialised Imaging) using a laser pulse, 30-µs duration, 640-nm wavelength (SI-LUX640, Specialised Imaging), for illumination. The high-speed camera comprises 16 CCDs that can be triggered independently to record up to 16 images with exposure times as short as 3 ns.

Impact speeds and rebound speeds were extracted from the image sequences. An example of impact of a 15-µm Al particle on an Al substrate is shown in Extended Data Fig. 7. The full-field video is available in the Supplementary Information (Video 3) and has a field of view of 637 × 478 µm. The measured impact speed is 765 m/s (±4%) and the rebound speed is 35 m/s (±4%). The coefficient of restitution is equal to 0.045 (±6%).

Before each impact test, particles to be ejected were selected using a secondary CCD camera. For each impact, the particle diameter was extracted from the image sequence. The measured particle diameters for Al, Ni, Cu and Zn are 14 ± 2, 10 ± 2, 14 ± 2 and 11 ± 2 µm, respectively. We have also conducted impact experiments on larger Al particles (45 µm) to resolve the particle deformation during impact with 10 ns time intervals.

**Finite Element Model.** The induced pressure in supersonic impact of metallic particles can greatly exceed the yield stress of the material. In such conditions, where the fractional deviations from stress isotropy are small, the behavior of the solid can be approximated by that of a fluid, and an Eulerian formulation has proven more effective than a Lagrangian one. Thus, we simulated particle behavior in an Eulerian domain consisting of fixed elements in space through which the material flows.

Using ABAQUS 6.14-3[30], we developed a three dimensional coupled thermo-mechanical dynamic explicit model to simulate the high-velocity-impact behavior of metallic particles. Extended Data Fig. 8 shows the stationary cube of $d \times d \times 2d$, where $d$ is the particle diameter, representing the possible positions of the particle material's flow. A scalar parameter, namely volume fraction, was assigned to each element such that the model represents a quarter-sphere at the beginning of the analysis (visible in the mesh in Extended Data Fig. 8). As the material flows through the Eulerian elements, the local volume fraction is computed and the material boundary is updated.

8-node thermally coupled linear Eulerian brick elements with reduced integration and hourglass control were used to discretize the Eulerian domain. After trial runs we chose the element size to be $d/50$. We considered the substrate to be a quarter cylinder of the same material as the particle, with $4d$ radius and $4d$ height. We discretized the substrate using Lagrangian elements and used a frictionless contact algorithm in the impact region. We started the simulations with an initial velocity of 200 m/s and increased the velocity in steps of 50 to 100 m/s until we observed impact-induced instabilities. We then refined our estimation of critical velocity with 25 m/s steps. Normal velocities were set to zero for all six faces of the Eulerian domain to prevent material loss. We applied a symmetry boundary condition to the substrate lateral faces (normal displacement was set to zero), and constrained the substrate bottom against all degrees of freedom. The initial temperature of the particle and the substrate was 298 K in all the simulations.

We simulated the impact behavior for four materials (Al, Cu, Ni and Ta) to cover a wide range of physical, thermal, and mechanical properties toward a unified description of critical velocity. In each case, the particle material is matched with the substrate material. To capture the hydrodynamic behavior of the particle upon impact the Mie-Grüneisen equation of state (equation 1) was used[27]. It defines the pressure, $P$, as a function of density and internal energy per unit mass $E_m$, with $\eta = 1 - \rho_0/\rho$ being the nominal volumetric compressive strain and $\Gamma_0$ a material constant referred to as the Grüneisen parameter. The Mie-Grüneisen equation is linear in energy and assumes a linear relationship between the shock velocity ($V_s$) and the particle velocity ($V_p$) (equation 2) where $C_0$ is the bulk speed of sound in the material and $s$ is an empirically determined material constant.

$$P = \frac{\rho_0 C_0^2 \eta}{(1 - s\eta)^2}\left(1 - \frac{\Gamma_0 \eta}{2}\right) + \Gamma_0 \rho_0 E_m \tag{1}$$

$$V_s = C_0 + sV_p \tag{2}$$

The dependency of the yield stress, $Y$, with plastic strain, strain rate, and temperature is reflected by the Johnson-Cook constitutive equation (equation 3)[26]. Here, $\varepsilon_p$ is the equivalent plastic strain, $\dot{\varepsilon}_p$ and $\dot{\varepsilon}_0$ are the applied and reference strain rates, $T_{ref}$, $T_{melt}$ are the reference and melting temperatures, $A$ is the initial yield stress of the material, $B$ and $n$ are the hardening coefficient and the exponent, respectively, $C$ and $m$ are constants describing the flow stress sensitivity to strain rate and temperature. Extended Data Table 1 summarizes the physical, thermal, and mechanical parameters[31–34] for the four materials used in the simulations. The Taylor-Quinney coefficient to dissipate plastic work as heat was considered to be 90% in our thermo-mechanical simulations.

$$Y = \left[A + B\varepsilon_p^n\right]\left[1 + C ln\frac{\dot{\varepsilon}_p}{\dot{\varepsilon}_0}\right]\left[1 - \left(\frac{T - T_{ref}}{T_{melt} - T_{ref}}\right)^m\right] \tag{3}$$

**REFERENCES**

1.    Richardson, J. E., Melosh, H. J. & Greenberg, R. Impact-induced seismic activity on asteroid 433 Eros: A surface modification process. *Science* **306,** 1526–1529 (2004).

2.    Chapman, C. R. & Morrison, D. Impacts on the Earth by asteroids and comets: assessing

the hazard. *Nature* **367,** 33–40 (1994).

3.     Malin, M. C., Edgett, K. S., Posiolova, L. V, McColley, S. M. & Dobrea, E. Z. N. Present-Day Impact Cratering Rate and Contemporary Gully Activity on Mars. *Science* **314,** 1573–1577 (2006).

4.     Knudson, M. D., Desjarlais, M. P. & Dolan, D. H. Shock-Wave Exploration of the High-Pressure Phases of Carbon. *Science* **322,** 1822–1825 (2008).

5.     Meyers, M. A. & Taylor Aimone, C. Dynamic fracture (spalling) of metals. *Prog. Mater. Sci.* **28,** 1–96 (1983).

6.     Voevodin, A. A., Bantle, R. & Matthews, A. Dynamic impact wear of TiCxNy and Ti-DLC composite coatings. *Wear* **185,** 151–157 (1995).

7.     Backman, M. E. & Goldsmith, W. The mechanics of penetration of projectiles into targets. *Int. J. Eng. Sci.* **16,** 1–99 (1978).

8.     Assadi, H., Gärtner, F., Stoltenhoff, T. & Kreye, H. Bonding mechanism in cold gas spraying. *Acta Mater.* **51,** 4379–4394 (2003).

9.     Dykhuizen, R. C. *et al.* Impact of High Velocity Cold Spray Particles. *J. Therm. Spray Technol.* **8,** 559–564 (1999).

10.    Visser, C. W. *et al.* Toward 3D Printing of Pure Metals by Laser-Induced Forward Transfer. *Adv. Mater.* **27,** 4087–4092 (2015).

11.    Park, M. & Schuh, C. A. Accelerated sintering in phase-separating nanostructured alloys. *Nat. Commun.* **6,** 6858 (2015).

12.    Moridi, A., Hassani-Gangaraj, S. M., Guagliano, M. & Dao, M. Cold spray coating: review of material systems and future perspectives. *Surf. Eng.* **30,** 369–395 (2014).

13.    Schmidt, T., Gärtner, F., Assadi, H. & Kreye, H. Development of a generalized parameter window for cold spray deposition. *Acta Mater.* **54,** 729–742 (2006).

14. Grujicic, M., Zhao, C. L., DeRosset, W. S. & Helfritch, D. Adiabatic shear instability based mechanism for particles/substrate bonding in the cold-gas dynamic-spray process. *Mater. Des.* **25,** 681–688 (2004).

15. Wu, J., Fang, H., Yoon, S., Kim, H. & Lee, C. The rebound phenomenon in kinetic spraying deposition. *Scr. Mater.* **54,** 665–669 (2006).

16. Schmidt, T. *et al.* From Particle Acceleration to Impact and Bonding in Cold Spraying. *J. Therm. Spray Technol.* **18,** 794–808 (2009).

17. Bae, G. *et al.* Bonding features and associated mechanisms in kinetic sprayed titanium coatings. *Acta Mater.* **57,** 5654–5666 (2009).

18. Grujicic, M., Saylor, J. R., Beasley, D. E., DeRosset, W. S. & Helfritch, D. Computational analysis of the interfacial bonding between feed-powder particles and the substrate in the cold-gas dynamic-spray process. *Appl. Surf. Sci.* **219,** 211–227 (2003).

19. Ko, K. H., Choi, J. O. & Lee, H. The interfacial restructuring to amorphous: A new adhesion mechanism of cold-sprayed coatings. *Mater. Lett.* **175,** 13–15 (2016).

20. Li, W.-Y., Li, C.-J. & Liao, H. Significant influence of particle surface oxidation on deposition efficiency, interface microstructure and adhesive strength of cold-sprayed copper coatings. *Appl. Surf. Sci.* **256,** 4953–4958 (2010).

21. Lee, J.-H. *et al.* High strain rate deformation of layered nanocomposites. *Nat. Commun.* **3,** 1164 (2012).

22. Lee, J.-H., Loya, P. E., Lou, J. & Thomas, E. L. Dynamic mechanical behavior of multilayer graphene via supersonic projectile penetration. *Science* **346,** 1092–1096 (2014).

23. Veysset, D. *et al.* Dynamics of supersonic microparticle impact on elastomers revealed by real–time multi–frame imaging. *Sci. Rep.* **6,** 25577 (2016).

24. Crossland, B. *Explosive welding of metals and its application*. (Oxford: Clarendon Press, 1982).

25.  Zel'dovich, Y. & Raiser, Y. *Physics of shock waves and high temperature hydrodynamics phenomena*. (New York: Academic, 1966).

26.  Johnson, G. R. & Cook, W. H. Fracture characteristics of three metals subjected to various strains, strain rates, temperatures and pressures. *Eng. Fract. Mech.* **21,** 31–48 (1985).

27.  Meyers, M. A. *Dynamic Behavior of Materials*. (Wiley, 1994).

28.  Field, J. E., Dear, J. P. & Ogren, J. E. The effects of target compliance on liquid drop impact. *J. Appl. Phys.* **65,** 533 (1989).

29.  Rein, M. Phenomena of liquid drop impact on solid and liquid surfaces. *Fluid Dyn. Res.* **12,** 61–93 (1993).

30.  *Dassault Systemes, ABAQUS 6.14-3 User's manual*. (2014).

31.  Meyers, M. A. *Dynamic Behavior of Materials*. (John Wiley & Sons, 1994).

32.  Kittel, C. *Introduction to Solid State Physics*. (John Wiley & Sons, 2004).

33.  Bae, G., Xiong, Y., Kumar, S., Kang, K. & Lee, C. General aspects of interface bonding in kinetic sprayed coatings. *Acta Mater.* **56,** 4858–4868 (2008).

34.  Murr, L. E. *et al.* Shock-induced deformation twinning in tantalum. *Acta Mater.* **45,** 157–175 (1997).

**ACKNOWLEDGEMENTS**

This work was supported by the U.S. Army through the Institute for Soldier Nanotechnologies. Funding was provided (in part) by the Army Research laboratory (Materials Manufacturing Technology Branch: RDRL-WMM-D) and by the U.S. Army Research Office, under Grant W911NF-13-D-0001. Support was also provided through Office of Naval Research DURIP Grant No. N00014-13-1-0676. MHG and DV thank Steve Kooi, Alex Maznev and Cyril Williams for valuable discussions. DV thanks Dmitro Martynowych for assistance in launching pad preparation. MHG thanks Atieh Moridi, Mario Guagliano and Richard Becker for valuable discussions.

**AUTHOR CONTRIBUTIONS**

All authors contributed to designing the research. D.V. and M.H.-G. performed the experiments. M.H.-G. performed the simulations. All authors analyzed the data and co-wrote the manuscript.

**COMPETING FINANCIAL INTERESTS**

The authors declare no competing financial interests.

**ADDITIONAL INFORMATION**

Correspondence and requests for materials should be addressed to CS.

**Supplementary Note 1: Impact-Induced Pressure**

For matched particle and substrate materials, the impact induced pressure in the particle and the substrate is given by

$$P = \frac{1}{2}(\rho C_0 V_i + s \frac{\rho V_i^2}{2}) \qquad (1)$$

where $P$ is the initial pressure in the shocked material, $\rho$ is the density of the particle/substrate, $C_0$ is the bulk speed of sound in the material, $V_i$ is the impact velocity, and $s$ is a material dependent coefficient (in the range of 1-2) that relates the velocity of the shock wave to the material velocity.

As discussed in the text, jetting is a pressure-governed mechanism. Identical pressures are induced in the particle and the substrate in matched materials impact, making both likely susceptible to jetting above the critical velocity, as demonstrated by the post-mortem SEM image in Extended Data Fig. 1.

The substrate in our models has been simulated in a Lagrangian domain where material deformation is represented by element deformation. Jetting in the Lagrangian domain causes excessive distortions of elements, inducing a high level of uncertainty in the calculated stresses, strains and temperatures. Accordingly, to avoid such numerical artifacts we effectively modeled the substrate as a deformation platform for the particle, and focused on capturing realistic jetting behavior in the particle, which is simulated in an Eulerian domain.

**Supplementary Note 2: Expanding the Relation between Critical Velocity and Bulk Speed of Sound to Include Particle Temperature and Size Effects**

In Extended Data Fig. 5, 6 we show that increasing particle temperature decreases the critical velocity in a manner that follows a square root dependency on the dimensionless temperature, $T^* = (T-T_{room})/(T_{melt}-T_{room})$ (Extended Data Fig. 5), and that increasing particle size results in higher temperatures in the jet region, which in turn affects the critical velocity. Variation of the maximum temperature in the jet region with the particle size reveals a power relation for all the examined metals with the exponents averaged to 0.12 +/- 0.01. We can therefore, incorporate these two effects into the relation between critical velocity and bulk speed of sound and semi-empirically expand it into equation 2 to address different metals, particle sizes and temperatures. In this equation, $n$ is the size effect exponent and $d_0$ is our reference size, i.e. 10 µm.

$$V_{cr} \approx 0.15 \times \sqrt{(\frac{d}{d_0})^{-n} \times (1 - T^*) \times \frac{B}{\rho}} \qquad (2)$$

While the temperature dependency in our equation conforms well with what has been previously put forward in the literature, we note that the size effect power-law exponents are much less (~3 times) than what has been measured earlier[13]. Such discrepancy can possibly be attributed to the fact that the earlier experiments used high-temperature gas driven setups, which are not capable of accurately isolating the size effect; smaller and larger particles would heat up to different temperatures and experience different turbulence effects in the nozzle during spraying. We note that the power value we measured in our experiments agrees well with our modeling results, where particle size is controlled to be the only variable.

**Supplementary Videos**

**Video 1 |** Impact of a 45-µm Al particle on an Al substrate at 605 m/s (± 3%) velocity. The video has a $637 \times 478$ µm field of view and an 820 ns duration. The particle does not adhere despite clear flattening and considerable deformation.

**Video 2 |** Impact of a 45-µm Al particle on an Al substrate at 805 m/s (± 3%) velocity. The video has a $637 \times 478$ µm field of view and a 515 ns duration. The particle does not rebound, but adheres to the substrate. Formation of an interfacial jet and plastic ejection of material is critical to supersonic adhesion.

**Video 3|** Impact of a 15-µm Al particle on an Al substrate at 765 m/s (±4%) velocity and its rebound. The video has a $637 \times 478$ µm field of view and a 1523 ns duration. The video is a montage of 16 images that have been used to measure the coefficient of restitution reported in Fig. 1C.

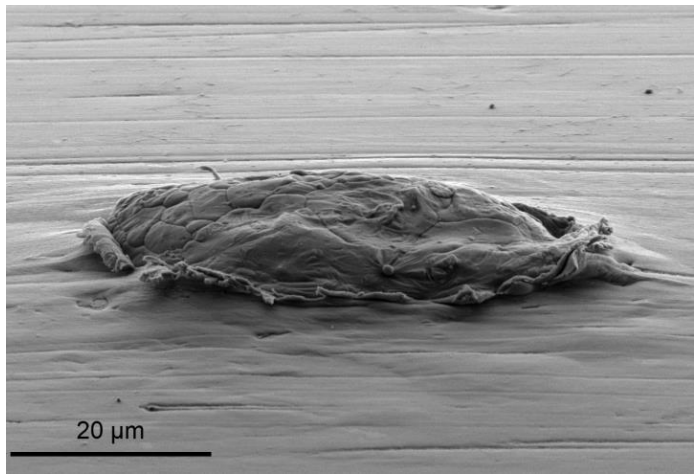

**Extended Data Figure 1 | Adhered Al Particle.** SEM image of an initially 30 µm spherical Al particle impacted and adhered to an Al substrate at 950 m/s. Material jet formation is evident at the periphery of the particle and the indentation. The authors are not aware of any prior work where the impact velocity of such an adhered particle has been directly measured prior to impact and reported.

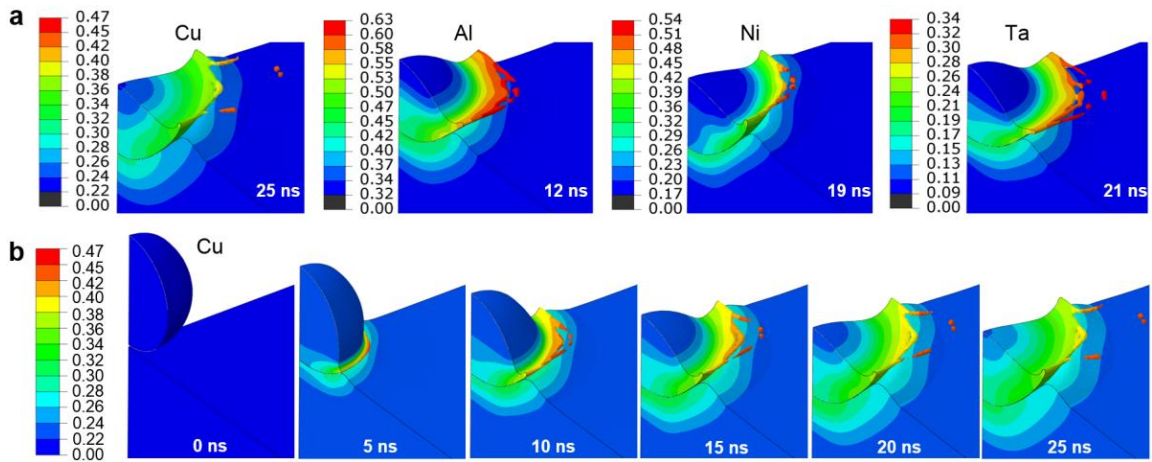

**Extended Data Figure 2 | Temperature distribution. a.** Homologous temperatures at the instant of maximum penetration at critical velocity are far less than the melting temperature in every case, revealing that jetting does not rely on softening associated with melting. **b.** These snapshots show the typical temperature distribution over the contact time, until maximum penetration, in Cu particle and substrate at critical velocity. No sign of melting is observed when the jet starts to form.

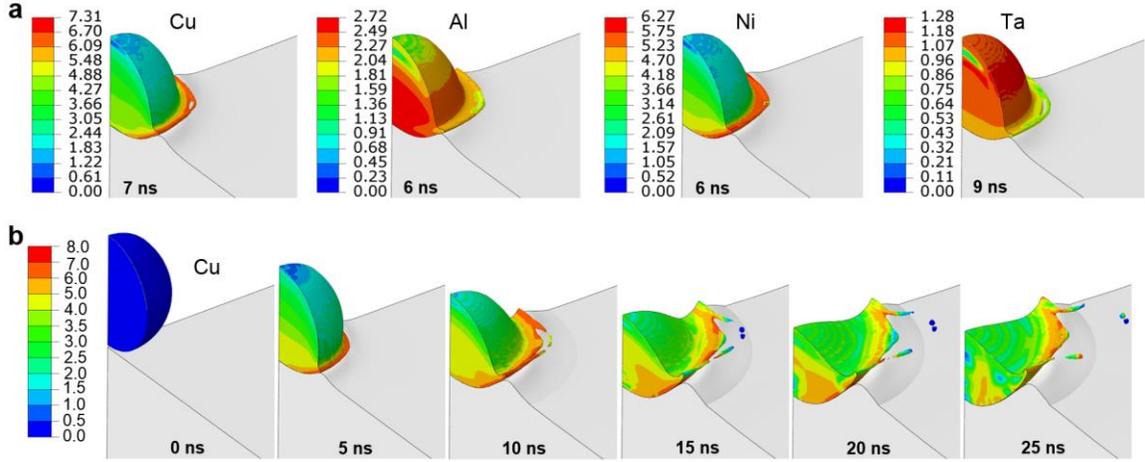

**Extended Data Figure 3 | Stress distribution. a.** Instantaneous yield stress normalized with the initial yield stress for the particles impacted at critical velocity, when material disintegration first begins in an already-formed jet. The yield stress in the jet region is higher than the initial yield stress for Cu, Al, Ni, and a large fraction of it for Ta. In no case has the yield strength been compromised to a value near zero. The fact that we observe an early material jet without indications of thermal softening suggest that adiabatic localization cannot be the cause for material jet formation. **b.** These snapshots show the instantaneous yield stress distribution normalized with the initial yield stress over the contact time, until maximum penetration, in a Cu particle at critical velocity. No sudden drop in stress is observed when the jet starts to form.

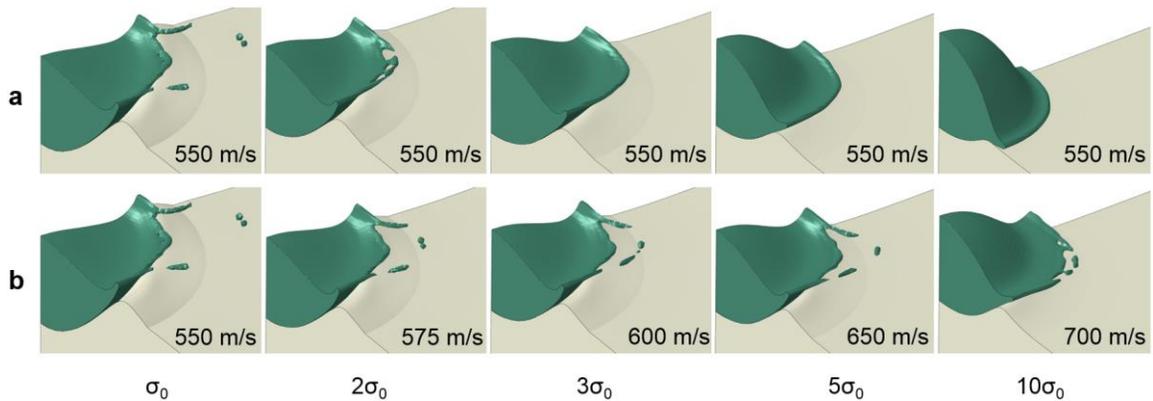

**Extended Data Figure 4 | Strength effect. a,** Simulated deformation of a 10-micron Cu particle impacting a Cu substrate at 550 m/s, where we increased the strength term in our models from 1 to 2,3,5 and 10 times higher values from left to right. We notice that material ejection and jet formation persists clearly despite a factor of 5 increase in strength, with even some tendency still observable at a factor of 10 increase in strength; this reveals the negligible role of strength in jet formation, and is a reasonable result in light of how far beyond the yield strength the pressures that evolve are for supersonic impacts (cf. Figure 3). **b,** The series of impact velocities at which we first observe fragmenting along with jetting in our simulations, as a function of the strength term in Cu. Note that an increase in material strength by a full order of magnitude only shifts the fragmentation velocity by ~25%. What is more, such strength levels ($10\sigma_o$ ~ 1 GPa) are unphysically high for Cu and constitute a significant fraction of the induced impact pressure, ~10 GPa. The strength effect seen here is thus extremely small and negligible over a physically reasonable range of material strengths, in contrast to some literature suggestions [8,13].

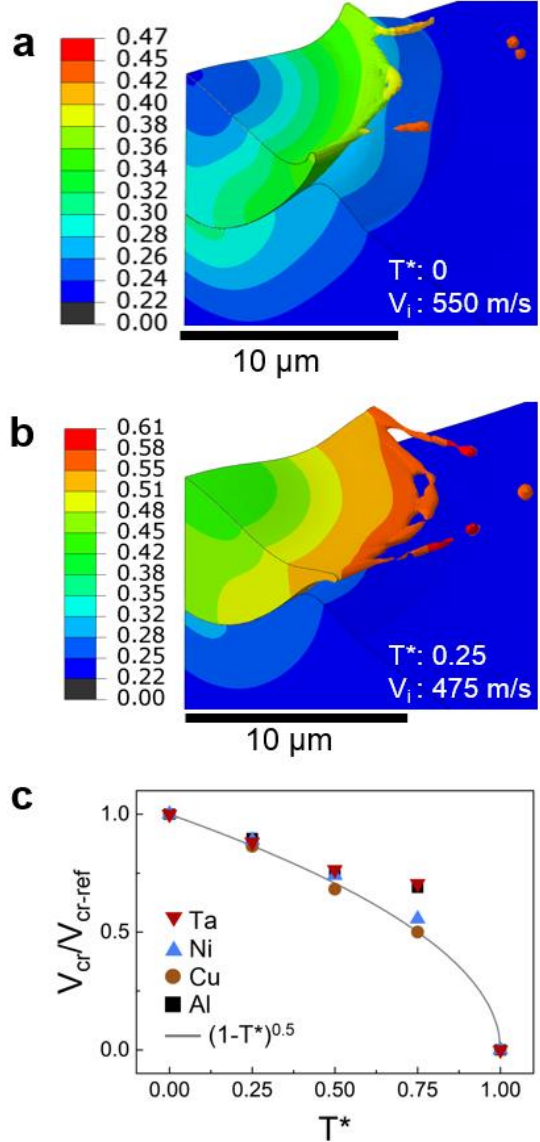

**Extended Data Figure 5 | Temperature effect.** We compare the deformation of 10-µm Cu particles with initial temperatures of **a,** 298 and **b,** 563 K (corresponding to T*=(T-T$_{room}$)/(T$_{melt}$-T$_{room}$) equal to 0 and 0.25) impacting a Cu substrate with the corresponding critical velocities i.e. 550 and 475 m/s respectively. The particle with higher initial temperature endures much more deformation and shows jet formation at a lower impact velocity. The particle at higher temperature penetrates 1 micron less into the substrate, and becomes flatter despite impacting at a lower velocity. **c,** Increasing initial particle temperature decreases the critical velocity in a manner that can be reasonably fitted with a square root relation that disappears at the melting point.

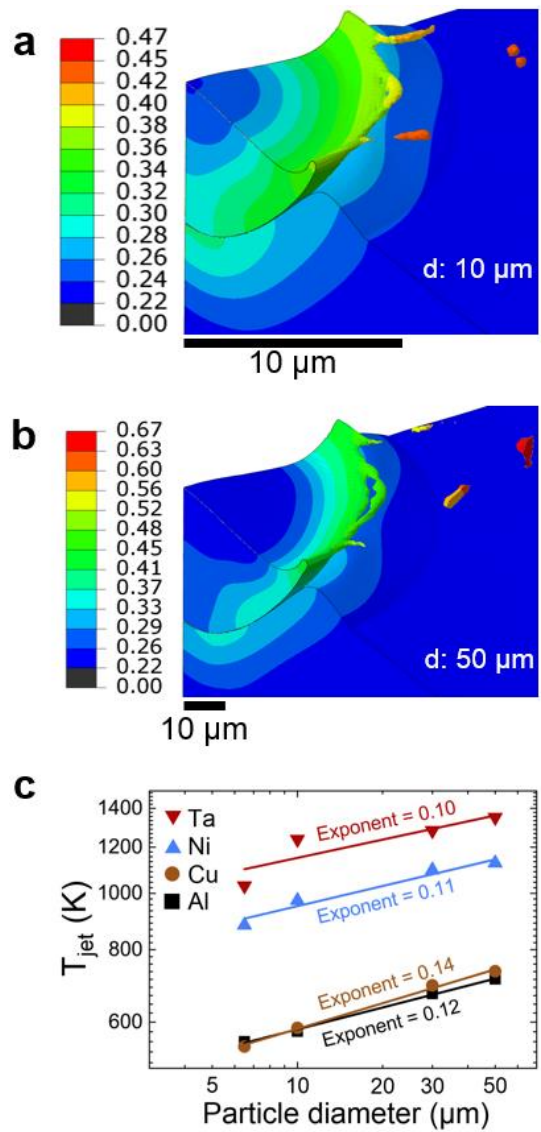

**Extended Data Figure 6 | Size effect.** We compare **a,** 10- and **b,** 50-micron Cu particles with the same initial temperature impacting Cu substrate with the same velocity when they are at the maximum penetration into the substrate. Although the general deformation features including the flattening ratio (particle height divided by particle diameter after deformation) are similar, the homologous temperature distribution reveals that the temperature in the jet region is higher for the larger particle and shows a larger spatial gradient. This makes larger particles softer and more prone to jetting. **c,** Increasing particle size increases the induced temperature in the particle jet with an apparent power law relation fitted as shown.

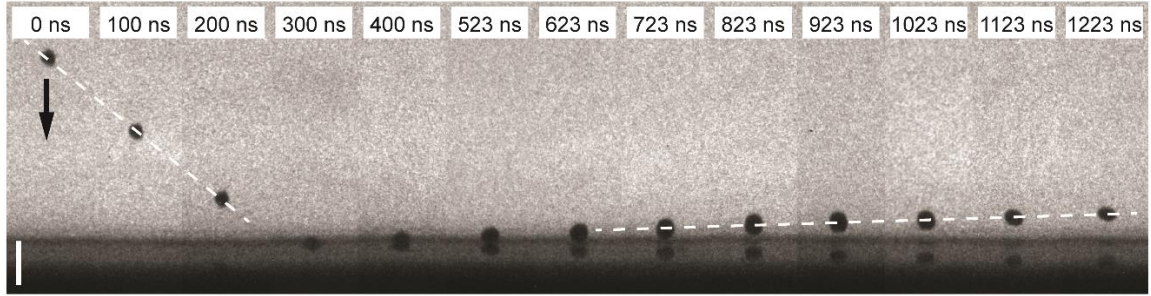

**Extended Data Figure 7** | Multi-frame sequences with 5 ns exposure time showing a single Al particle impacting on an Al substrate. The microparticle arrives from the top of the field of view with a speed of 765 m/s, impacts the substrate and subsequently rebounds with a speed of 35 m/s. The relative delay from the initial image is shown at the top of each frame. The images are cropped from their original size to show the region of interest (see Supplementary Videos S1 for a full-field view). The vertical scale bar is 50 µm.

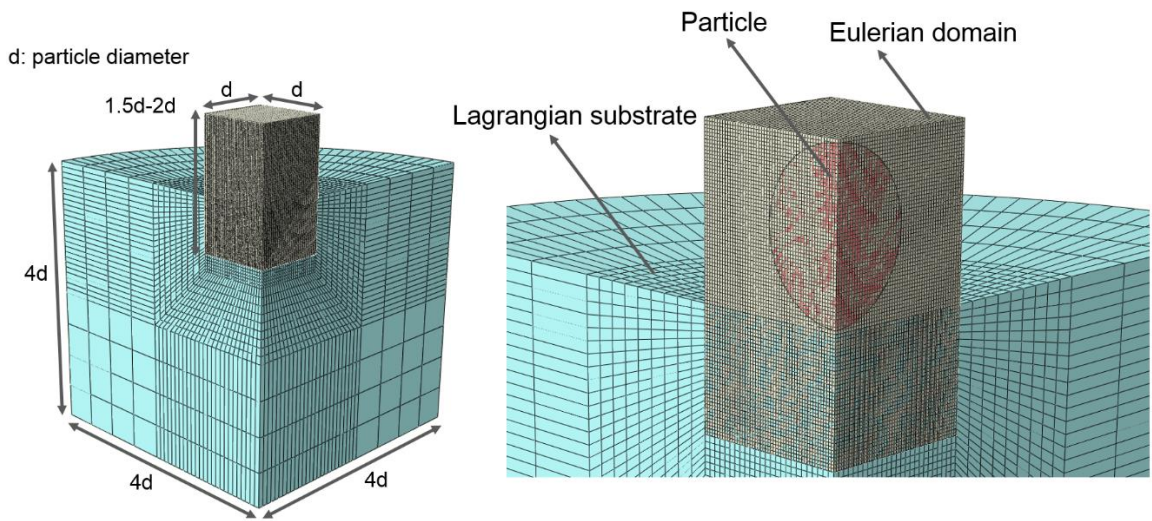

**Extended Data Figure 8 |** Finite element model of a metallic particle impacting on a metallic substrate. The particle is treated in the Eulerian domain where material flows through the elements. The substrate is treated in the Lagrangian domain where material deformation is captured by element deformation.

**Extended Data Table 1 |** Physical, thermal and mechanical parameters of the four materials used in the simulation[31-34].

|  | **Cu** | **Ni** | **Al** | **Ta** |
|---|---|---|---|---|
| Density (kg/m$^3$) | 8960 | 8908 | 2700 | 16690 |
| Specific heat (J/kg K) | 384.6 | 444.2 | 896.9 | 140.2 |
| Melting temperature (K) | 1357 | 1728 | 933 | 3290 |
| Heat of Fusion (kJ/kg) | 208.7 | 297.8 | 396.9 | 202.1 |
| Conductivity (W/m K) | 401 | 90.9 | 237 | 57.5 |
| Shear Modulus (GPa) | 48 | 76 | 26 | 69 |
| Poisson's ratio | 0.34 | 0.31 | 0.35 | 0.34 |
| Bulk Modulus (GPa) | 140 | 180 | 76 | 200 |
| $C_0$ (m/s) | 3952.8 | 4495.2 | 5305.5 | 3461.7 |
| s | 1.49 | 1.44 | 1.339 | 1.2 |
| $\Gamma_0$ | 2.01 | 1.83 | 2.17 | 1.61 |
| A (MPa) | 90 | 163 | 148.4 | 684.5 |
| B (MPa) | 292 | 648 | 345.5 | 205.3 |
| n | 0.31 | 0.33 | 0.183 | 0.78 |
| C | 0.025 | 0.006 | 0.001 | 0.043 |
| m | 1.09 | 1.44 | 0.895 | 0.344 |
| $\dot{\varepsilon}_0$ (s$^{-1}$) | 1 | 1 | 1 | 3500 |
| $T_{ref}$ (K) | 298 | 298 | 293 | 298 |